 \documentclass[twocolumn,epjc3]{svjour3}

\journalname{Eur. Phys. J. A}
 
\usepackage{epsfig}
\usepackage{color}
\usepackage{float}
\usepackage{amssymb}
\usepackage{amsmath, bbm}
\usepackage[figuresright]{rotating}

\usepackage[numbers,sort&compress]{natbib}
\usepackage{hyperref}

\newlength{\feynwidth} 
\usepackage{amssymb}
\usepackage{amsmath, bbm}
\usepackage[figuresright]{rotating}
\usepackage{capt-of}
\usepackage{multirow}

\def\nl{\nonumber\\}

\newcommand{\La}{{\Lambda}}

\newcommand{\be}{\begin{eqnarray}}
\newcommand{\ee}{\end{eqnarray}}

\newcommand{\eeb}{{e^+e^-}}
\newcommand{\LLb}{{\Lambda\bar\Lambda}}
\newcommand{\pLb}{{p\bar\Lambda}}
\newcommand{\nLb}{{n\bar\Lambda}}
\newcommand{\Lpb}{{\bar p \Lambda}}

\newcommand{\BBb}{{B \bar B}}

\newcommand{\ppbar}{{p \bar p}}
\newcommand{\nnbar}{{n \bar n}}
\begin{document}

\title{{\boldmath$p\bar\Lambda$} final-state interaction in the reactions 
{\boldmath$e^+e^- \to K^- p \bar \Lambda$} and {\boldmath$J/\psi \to K^- p \bar \Lambda$}}
\titlerunning{$ p\bar\Lambda$ final-state interaction}  

\author{Johann Haidenbauer \thanksref{addr1,e2}
\and Ulf-G. Mei{\ss}ner \thanksref{addr2,addr1,e3}
}
\thankstext{e2}{e-mail: j.haidenbauer@fz-juelich.de}
\thankstext{e3}{e-mail: meissner@hiskp.uni-bonn.de}

\institute{Institute for Advanced Simulation (IAS-4), 
Forschungszentrum J\"ulich, D-52428 J\"ulich, Germany
  \label{addr1}
           \and
           Helmholtz-Institut f\"ur Strahlen- und Kernphysik 
           and Bethe Center for Theoretical Physics,
           Universit\"at Bonn, D-53115 Bonn, Germany \label{addr2}
}

\date{\today}

\maketitle

\abstract{
Near-threshold
$p\bar\Lambda$ mass spectra for the reactions $e^+e^- \to K^- p\bar\Lambda$ 
and $J/\psi \to K^- p\bar\Lambda$ are investigated with an emphasis on the 
role played by the interaction in the $p\bar\Lambda$ system. 
As guideline for the $p\bar\Lambda$ interaction a variety 
of $\Lambda\bar\Lambda$ potential models is considered that have been established 
in the analysis of data on $p\bar p\to \Lambda\bar\Lambda$ in the past.
Arguments why the properties of the $p\bar\Lambda$ and $\Lambda\bar\Lambda$ 
interactions can be expected to be very similar are provided. 
It is shown that the near-threshold enhancement in the invariant mass observed 
for the $e^+e^-$ reaction can be reproduced quantitatively by the assumed 
$p\bar\Lambda$ final-state interaction in the partial wave suggested by an  
amplitude analysis of the experiment. 
The effect of the $p\bar\Lambda$ final-state interaction in other 
decays is explored, including the recently measured reactions
$B^- \to J/\psi\, \bar p \Lambda$ and $B^+ \to J/\psi\, p \bar\Lambda$. 
It is found that the final-state interaction improves the description of 
the measured invariant mass near threshold in most cases. 

}

\section{Introduction} 
One of the phenomena that caught a wider attention over the last two decades is the
near-threshold enhancement in the invariant mass of baryon-antibaryon
($\BBb$) systems observed in various heavy-meson decays and also in $e^+e^-$ 
collisions. The most spectacular example is definitely the anomalously strong 
enhancement detected in the $\ppbar$ spectrum in the reaction 
$J/\psi \to \gamma \ppbar$ \cite{BES:2003aic,CLEO:2010fre,BESIII:2011aa}. 
Less prominent but still noticeable enhancements as compared to the 
phase-space behavior have been also observed in other decays like
$J/\psi \to \pi^0 \ppbar$ \cite{BES:2003aic,BES:2009ufh}, 
$\psi' \to \gamma \ppbar$ \cite{BESIII:2011aa,CLEO:2010fre}, 
and in 
$\eeb \to \ppbar$ \cite{BaBar:2013,CMD-3:2019}. 
Reactions with a $\LLb$ system in the final state have been studied, too
\cite{BESIII:2013Psi,BESIII:2022Psi,
Aubert:2007,BESIII:2018,BESIII:2023ee,
BESIII:2021,BESIII:2022}
and in this case clear evidence for a near-threshold enhancement has
been observed in reactions like $\eeb \to \eta\LLb$ \cite{BESIII:2022}
and $\eeb \to \phi\LLb$ \cite{BESIII:2021}.
The most puzzling result is definitely the large non-zero cross section 
barely $1$~MeV away from the $\LLb$ threshold, reported in a measurement of 
the reaction $\eeb \to \LLb$ by the BES\-III Collaboration~\cite{BESIII:2018}. 

The two most striking observations mentioned have triggered many studies 
offering a wide spectrum of possible reasons for the origin of the detected 
enhancement.
Those range from very exotic explanations like glueballs
\cite{Kochelev2005,Li2005,He2005},
or of evidence for new resonances or for $\BBb$ bound states  
\cite{BES:2003aic,BESIII:2011aa,Datta,MLYan1,MLYan2,Cao:2018,Li:2022}, 
to a more conventional interpretation in 
form of a final-state interaction (FSI) between the produced $\BBb$ pair
\cite{Kerbikov:2004,Sibirtsev:2005,Haidenbauer:2006, 
Loiseau:2005,Entem:2007,Loiseau:2009,Chen:2010,Kang:2015,Milstein:2017,Haidenbauer:2014,
Haidenbauer:2016,Haidenbauer:2021,Milstein:2021,Salnikov:2023,Haidenbauer:2023}. 

So far, much less attention was paid to the $\pLb$ interaction. 
With the present work we want to provide a remedy of this neglect.
Indeed, also for this $\BBb$ system there is a wealth of data available, 
though with somewhat less statistics as compared to $\ppbar$ 
and/or $\LLb$, and a near-threshold enhancement has been observed 
in some reactions.
To be concrete, there are measurements of the $\pLb$ ($\pLb$) invariant
mass spectrum near the threshold in the reactions 
$J/\psi \to K^-\pLb$ and $\psi (3686) \to K^-\pLb$
by the BES Collaboration \cite{BES:2004}, 
and for 
$\chi_{c0} \to K^+\bar p\La$ \cite{BESIII:2013},
$\psi (3686) \to K^{*+}\bar p\La$ \cite{BESIII:2019}, and
$\eeb\to K^-\pLb$ \cite{BESIII:2023}
by BESIII\footnote{
Note that in experiments with neutral initial states like $J/\psi$ or $\eeb$
usually the data for the two charge-conjugated decay modes are combined, 
e.g. those for $K^- \pLb$ and $K^+\Lpb$. 
Thus, throughout our paper we will not distinguish between $\pLb$ and $\Lpb$. 
}. 
From the enhancement seen in $J/\psi \to K^-\pLb$ the resonance parameters
$m= 2075\pm 12 \pm 5$~MeV, $\Gamma = 90\pm 35$~MeV (assuming an $S$-wave) 
or $m= 2044\pm 17$~MeV, $\Gamma = 20\pm 45$~MeV ($P$-wave)   
have been deduced \cite{BES:2004}. 
In case of $\eeb\to K^-\pLb$ an amplitude analysis has been performed and
the enhancement was attributed to the $1^+$ ($P$-wave) state. 
Here the reported parameters are 
$m= 2084^{+4}_{-2} \pm 9$~MeV, $\Gamma = 58^{+4}_{-3}\pm 25$~MeV 
\cite{BESIII:2023}.
A fit to the enhancement observed in $\chi_{c0} \to K^+\bar p\La$ 
yielded $m= 2053\pm 13$~MeV, $\Gamma = 292\pm 14$~MeV, with preference for $0^-$
($^1S_0$ state) \cite{BESIII:2013}.
However, in all aforementioned experiments there is no distinct peak-like
structure in the measured spectrum and the deduced resonances overlap
with the threshold (within their width) or even lie below the $\pLb$ threshold,
which is at $2053.95$~MeV. Thus, using a Breit-Wigner type ansatz is incorrect,
as the underlying assumption of a slowly varying background is certainly
not fulfilled in the vicinity of a threshold.
 
There are also measurements for $J / \psi \to  K^0_S n\bar \Lambda$ \cite{BES:2007} 
and the decays $\chi_{cJ}\to K^{0}_{S}n\bar\Lambda$ 
for $J=0,1,2$ \cite{BESIII:2021c}. 
In addition there are preliminary data by the
GlueX Collaboration on $\gamma p \to \La \bar\La p$
\cite{Li:2020,Pauli:2022}. 
Finally, measurements of the $\pLb$ ($\pLb$) invariant mass
have been reported for $B^0 \to \pi^- p \bar\La$ by the
Belle \cite{Belle:2003taw} and BaBar \cite{BaBar:2009ess}
Collaborations, and for the reactions
$B^- \to J/\psi\, \bar p \Lambda$ \cite{Belle:2005,LHCb:2023} and 
$B^+ \to J/\psi\, p \bar\Lambda$ \cite{CMS:2019}. 

In the present work we want to review and re-examine the information on 
the $\pLb$ interaction. In particular, we want to investigate to what extent 
the FSI between these two baryons plays a role in understanding and interpreting 
their invariant-mass spectum in the near-threshold region. 
The main focus will be certainly on the decay of $J/\psi$ and 
the signal in $e^+e^-$ collisions, where evidence for resonances
has been claimed as mentioned above. However, we will also explore
the situation for several other reactions where near-threshold results
for the $\pLb$ spectrum have been published. 

Without ready access to a suitable $\pLb$ potential, 
as guide line for the momentum dependence of the $\pLb$ FSI a 
variety of $\Lambda\bar\Lambda$ potential models is utilized that have 
been established in the analysis of data on $p\bar p\to \Lambda\bar\Lambda$ 
\cite{PS185} in the past \cite{Haidenbauer:1992,Haidenbauer:1992B}. 
Arguments why the properties of the $p\bar\Lambda$ and 
$\Lambda\bar\Lambda$ interactions can be expected to be very similar 
are provided below. 
One of those are measurements of two-particle correlation 
functions in heavy-ion collisions \cite{STAR:2006,Kisiel:2014} 
by the STAR Collaboration 
(in Au-Au collisions at $\sqrt{s_{NN}}=200$~GeV) 
and in high-energy $pp$ collisions
by the ALICE Collaboration \cite{ALICE:2020,ALICE:2022} 
which yielded very similar results for $\pLb$ and $\LLb$.

The paper is structured in the following way. In the subsequent section we
provide a brief summary of the employed formalism for treating the FSI effects. 
In Sect.~\ref{sec:Results} we present our results. Specifically, we investigate
the $\pLb$ FSI for $\eeb\to K^-\pLb$ and $J/\psi \to K^-\pLb$ and 
then show our predictions for the $\pLb$ invariant mass spectra measured in the 
reactions $B^- \to J/\psi\, \bar p \Lambda$ and $B^+ \to J/\psi\, p \bar\Lambda$. 
We also discuss the situation for $\psi (3868) \to K^-\pLb$ and 
$\chi_{cJ} \to K^-\pLb$ among others. 
The paper closes with a short summary. 


\section{Treatment of the $\BBb$ final-state interaction}
\label{Sec:Form}

How we treat the FSI is described and discussed in detail in 
Refs.~\cite{Sibirtsev:2005,Kang:2015,Haidenbauer:2023}. Thus, below we provide
only a summary of the essential formulae. 
The calculation of the $\BBb$ invariant-mass spectrum is based on the distorted 
wave Born approximation (DWBA), where the reaction amplitude $A$ is given 
schematically by \cite{Sibirtsev:2005,Kang:2015}
\begin{equation}
A=A^0 + A^0 G^{\BBb} T^{\BBb} \ . 
\label{eq:DWBA}
\end{equation}
Here, $A^0$ is the elementary (or primary) production amplitude, $G^{\BBb}$ the free $\BBb$ Green's
function, and $T^{\BBb}$ the $\BBb$ reaction amplitude.
The explicit expression for a specific partial wave with orbital angular momentum $L$ reads

\begin{eqnarray}
&&A_L(k)=\bar A^0_L k^L \times \nonumber \\ 
&&\times\left[1+ \int_0^\infty \frac{dp p^2}{(2\pi)^3} 
\frac{p^L}{k^L}\frac{1}{\sqrt{s}-E_p+i0^+}T_{L}(p,k;\sqrt{s})\right] ,\nl 
\label{eq:dwba2}
\end{eqnarray}
with $\sqrt{s}  = M_\BBb = \sqrt{m_{B_1}^2 + k^2} + \sqrt{m_{B_2}^2 + k^2}$ 
the $\BBb$ invariant mass (energy in the $\BBb$ subsystem)
and $E_p = \sqrt{m_{B_1}^2 + p^2} + \sqrt{m_{B_2}^2 + p^2}$.
The momentum factor $k^L$ is pulled out so that $\bar A^0_L$ is essentially a 
constant near threshold (see the detailed discussion in \cite{Haidenbauer:2023}). 
The $\BBb$ $T$-matrix, $T_{L}(p,k;\sqrt{s})$, is obtained by solving the 
Lipp\-mann-Schwinger (LS) equation,
\begin{eqnarray}
&&T_{L}(p',k;\sqrt{s})=V_{L}(p',k)+ \nl
&& \int_0^\infty \frac{dpp^2}{(2\pi)^3} \, V_{L}(p',p)
\frac{1}{\sqrt{s}-E_p+i0^+}T_{L}(p,k;\sqrt{s})~,\nl 
\label{LS}
\end{eqnarray}
for a specific $\BBb$ potential $V_L$. In the case of coupled partial waves like the $^3S_1$--$^3D_1$
the corresponding coupled LS equation is solved \cite{Haidenbauer:2014}, and then 
$T_{LL}$ is used in Eq.~(\ref{eq:dwba2}).

The $\pLb$ invariant-mass spectrum is calculated via 
\begin{equation}
\frac{d\sigma}{dM} \propto k\,|A_L(k)|^2 \ ,
\label{eq:INV}
\end{equation}
which is valid when the (total) energy is significantly larger than the $K^-\pLb$ 
threshold energy. Then for low $\pLb$ invariant masses the relative 
momentum of the third particle is large and its interaction does not distort the 
signal of interest. This condition is fulfilled by most of the reactions where 
measurements by the BESIII Collaboration are available.
In cases like $B^+ \to J/\psi \pLb$ where the available phase space is small
we evaluate the three-particle phase space explicitly \cite{Byckling}, 
\begin{equation}
\frac{d\sigma}{dM}\propto 
\lambda^{1/2}(s_{\rm tot},M^2,m_{K}^2)
\lambda^{1/2}(M^2,m_{p}^2,m_{\La}^2)\,
|A_L(k)|^2 ,
\label{eq:INV2}
\end{equation}
where the K\"all\'en function $\lambda$ is defined by
$\lambda (x,y,z)={((x-y-z)^2-4yz})/{4x}\,$,
$M \equiv M_{p\bar \La}$ is the invariant mass of the $p{\bar \La}$
system and $s_{\rm tot}$ is the total energy squared. 


The results shown below are based on the $\LLb$ potentials used in 
our analysis of the reactions $e^+e^- \to \phi \LLb$ and 
$e^+e^- \to \eta \LLb$ \cite{Haidenbauer:2023}.  
Indeed, there is an almost complete lack of $\pLb$ potentials \cite{Huang:2012}, 
among other reasons because there is no extensive empirical information 
that would allow to establish such an interaction, quite in contrast to $\LLb$ 
which can be reasonably well constrained by data on the
reaction $\ppbar \to \LLb$, for which differential cross sections, 
polarizations, etc. have been measured down to energies very close to 
the threshold~\cite{PS185}.
The employed $\LLb$ potentials (I-IV) are described in detail in
Refs.~\cite{Haidenbauer:1992,Haidenbauer:1992B}, see also the 
discussion in \cite{Haidenbauer:2023}. They consist of an 
elastic part generated by meson exchanges, and an annihilation 
part in form of a phenomenological optical potential.
As already said above, 
all of them have been determined in a fit to the wealth of
$\ppbar \to \LLb$ data collected by the P185 Collaboration~\cite{PS185}.

We decided to employ those $\LLb$ potentials for generating the FSI
effects because first exploratory calculations with them yielded quite 
promising results for the measured $\pLb$ spectra. Anyway, it can be 
expected that the interactions in the $\BBb$ systems exhibit some universal 
properties which are essential for reproducing the threshold 
behavior of the invariant-mass spectrum. The most important aspect 
is that annihilation processes dominate the $\LLb$ as well as 
the $\pLb$ interactions. Furthermore, in both cases one-pion exchange 
is not possible, and thus there is no long-range elastic contribution.
Finally, in both cases there is only one isospin state, so that no 
near-threshold coupling, say, like that between $\ppbar$ and $\nnbar$, exists. 
Taking those aspects together, it is a reasonable working hypothesis to 
assume that the properties of the FSI effects due to the $\LLb$ and $\pLb$ 
interactions are very similar. 

\begin{figure}
\begin{center}
\includegraphics[height=88mm,angle=-90]{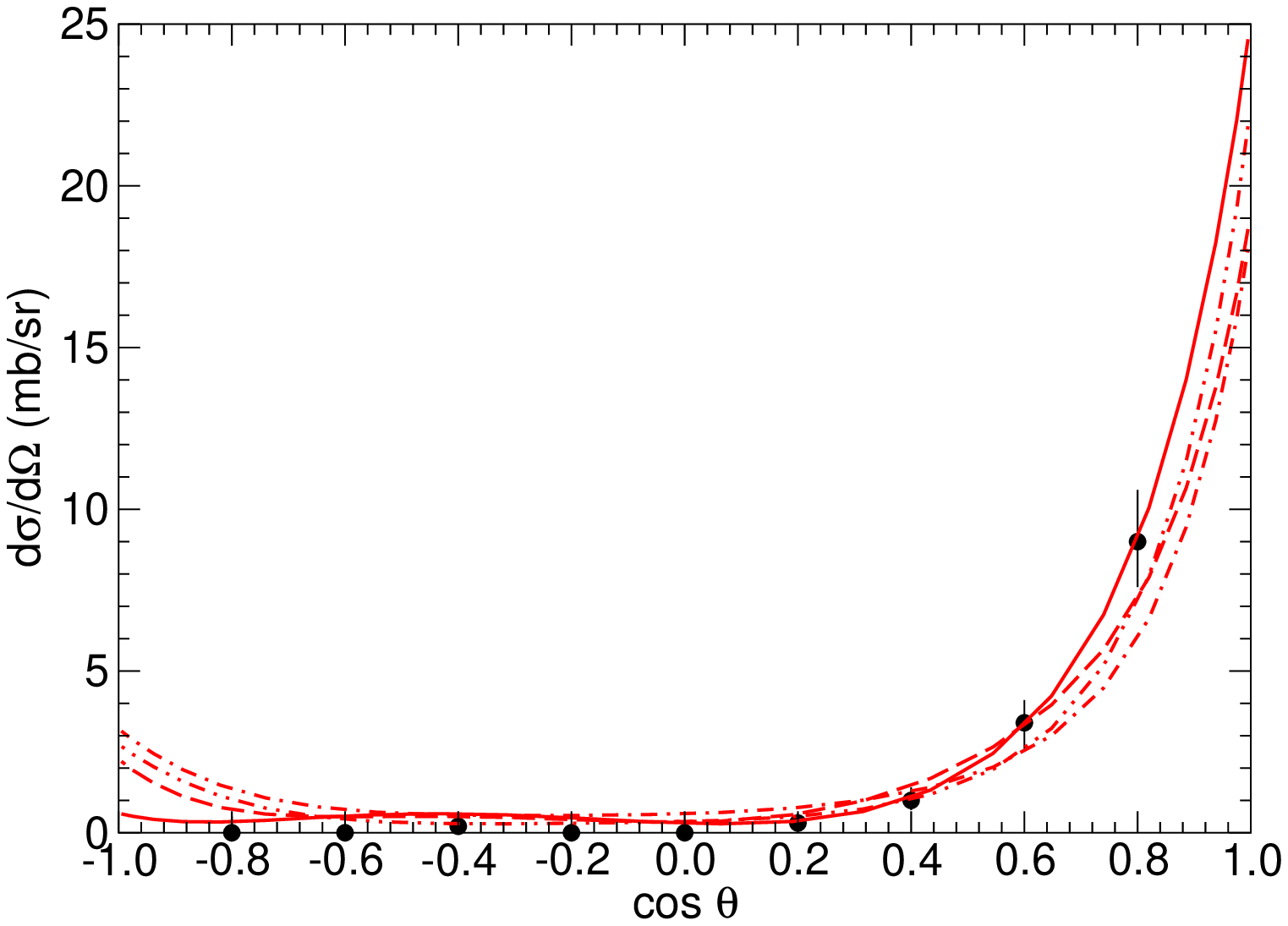}
\caption{Differential cross section for $\pLb$ scattering 
at $p_{lab}= 1.074\pm 0.017$~GeV \cite{BESIII:2024}. 
The curves are predictions by the $\LLb$ interactions I-IV, 
see Ref.~\cite{Haidenbauer:2023}, at $1.05$~GeV/c. 
}
\label{fig:XS}
\end{center}
\end{figure}

It should be added that 
the similarity of the $\LLb$ and $\pLb$ interactions is supported by 
measurements of the pertinent correlation functions. Available data from 
the STAR and ALICE Collaborations demonstrate that the momentum dependence 
established for the two systems is practically the same  
\cite{STAR:2006,Kisiel:2014,ALICE:2020,ALICE:2022}. 
Interestingly, even observables like the differential cross section
could be similar, as suggested by the comparison of a recent measurement
of $\pLb$ scattering at roughly $1$~GeV by BESIII~\cite{BESIII:2024}
with a prediction for $\LLb$ employing the phenomenological potentials
published by the J\"ulich~Group \cite{Haidenbauer:1992,Haidenbauer:1992B}, 
see Fig.~\ref{fig:XS}. Though, in principle, one cannot exclude that this
agreement is purely accidental, very likely it is an indication that 
both interactions are really dominated by strong and universal 
annihilation processes, as already argued in~\cite{BESIII:2024}. 

In principle, one could try to construct a phenomenological $\pLb$ potential.
But in view of the semi-quanti\-ta\-tive agreement of the $\LLb$ results 
with the $\pLb$ data (after adjusting for phase-space differences), 
to be reported below, one would anyway adopt 
the $\LLb$ potential as determined 
in the fit to the $\ppbar \to \LLb$ data as starting point. 
In fact, by recalling the Schr\"odinger equation, 
\begin{equation}
\nonumber
-u''(k,r) + 2 \mu_{B_1B_2} V_{B_1B_2}(r)\, u(k,r) = k^2\, u(k,r) \ ,
\end{equation}
with $u(k,r)$ the wave function and $\mu_{B_1B_2}$ the 
reduced mass, obtaining an $\pLb$ potential can be simply
realized by requiring that $\mu_{\pLb} V_{\pLb} \simeq \mu_{\LLb} V_{\LLb}$.
Then the $k$ dependence of all observables, including the mass
spectrum for the $\pLb$ and $\LLb$ systems would be identical.
The ratio of the reduced masses is $1.09$ so that the $\pLb$ 
potential established that way would be about $10$~\% stronger   
than that for $\LLb$. 

Since there are no $C$-parity restrictions in reactions
involving the $K$ meson, the selection rules for the
decay to the $K\pLb$ final state are less rigid and,
in general, both spin states of the $\pLb$ ($\Lpb$)
system can occur. An overview of the allowed partial waves is given in
Tab.~\ref{tab:JP}. 

\vskip 0.2cm
\begin{table}
\renewcommand{\arraystretch}{1.1}
\begin{center}
\caption{Allowed $\pLb$ partial waves ($^{2S+1}L_J$) and $J^{P}$ assignments 
(up to $P$-waves) for various initial states. $s,p,d$ indicate  the
relative orbital angular momentum of the $K$ meson. 
}
\begin{tabular}{|c|c|}
\hline
initial & partial waves \\
state   &               \\
\hline 
 $J/\psi [1^{-}] $ & ${}^1S_0p \, [0^{-}], \,{}^3S_1p \, [1^{-}], \, {}^{1,3}P_1s \, [1^{+}], \, {}^3P_2d \, [2^{+}]$\\
 $e^+e^- [1^{-}] $ & ${}^1S_0p \, [0^{-}], \,{}^3S_1p \, [1^{-}], \, {}^{1,3}P_1s \, [1^{+}], \, {}^3P_2d \, [2^{+}]$\\
$\chi_{c0} [0^{+}] $ & ${}^1S_0s \, [0^{-}], \, {}^{1,3}P_1p \, [1^{+}]$ \\
$\chi_{c1} [1^{+}] $ & ${}^3S_1s \, [1^{-}], \, {}^{1,3}P_1p \, [1^{+}], \, {}^3P_2p \, [2^{+}]$  \\
$\chi_{c2} [2^{+}] $ & ${}^1S_0d \, [0^{-}], \, {}^3S_1d \, [1^{-}], \, {}^{1,3}P_1p \, [1^{+}], \, {}^3P_2p \, [2^{+}]$  \\
\hline
\end{tabular}
\label{tab:JP}
\end{center}
\renewcommand{\arraystretch}{1.0}
\end{table}


\section{Results}
\label{sec:Results} 

In the following we explore the $\pLb$ inva\-riant-mass spectrum
as measured in various heavy-meson decays and in $\eeb$.
We start with the reactions $\eeb\to K^-\pLb$ and $J/\psi\to K^-\pLb$  
where data are available from the BES \cite{BES:2004} and BESIII \cite{BESIII:2023}
Collaborations and where evidence for a narrow structure near the $\pLb$ threshold
has been claimed \cite{BESIII:2023}. 
Then we take a look at $B^+\to J/\psi \pLb$ and $B^-\to J/\psi \Lpb$,
where the CMS \cite{CMS:2019} and LHCb \cite{LHCb:2023} Collaborations 
provided invariant-mass spectra with excellent momentum resolution and where in 
the latter measurement evidence for a pentaquark, the $P^\Lambda_{\psi s}(4338)$,
has been reported. 
Finally, we illustrate the situation for other reactions where results for 
the $\pLb$ or $\nLb$ invariant mass have been presented, although with lower
resolution. 
As mentioned above, the present investigation exploits the momentum dependence 
predicted by the $\LLb$ potentials from~\cite{Haidenbauer:1992,Haidenbauer:1992B} 
which has been already examined in the study of the electromagnetic form factors 
of the $\Lambda$ in the time-like region~\cite{Haidenbauer:2016}  
and in various meson decays with $\LLb$ in the final state~\cite{Haidenbauer:2023}. 
 
As a reminder, and
as already emphasized in Refs.~\cite{Kang:2015,Haidenbauer:2023} 
the validity of treating FSI effects via Eqs.~(\ref{eq:dwba2}) 
and (\ref{eq:INV}) is clearly limited, say to excess energies 
of $50$ to $100$~MeV. 
With increasing invariant mass the momentum dependence of the reaction/production 
mechanism should become more and more relevant and will likewise influence the 
invariant-mass spectrum. 

Note further that for none of the measurements considered below the data have
been published in numerical form. Thus, we digitized them from the pertinent
figures. 

\begin{figure}
\begin{center}
\includegraphics[height=88mm,angle=-90]{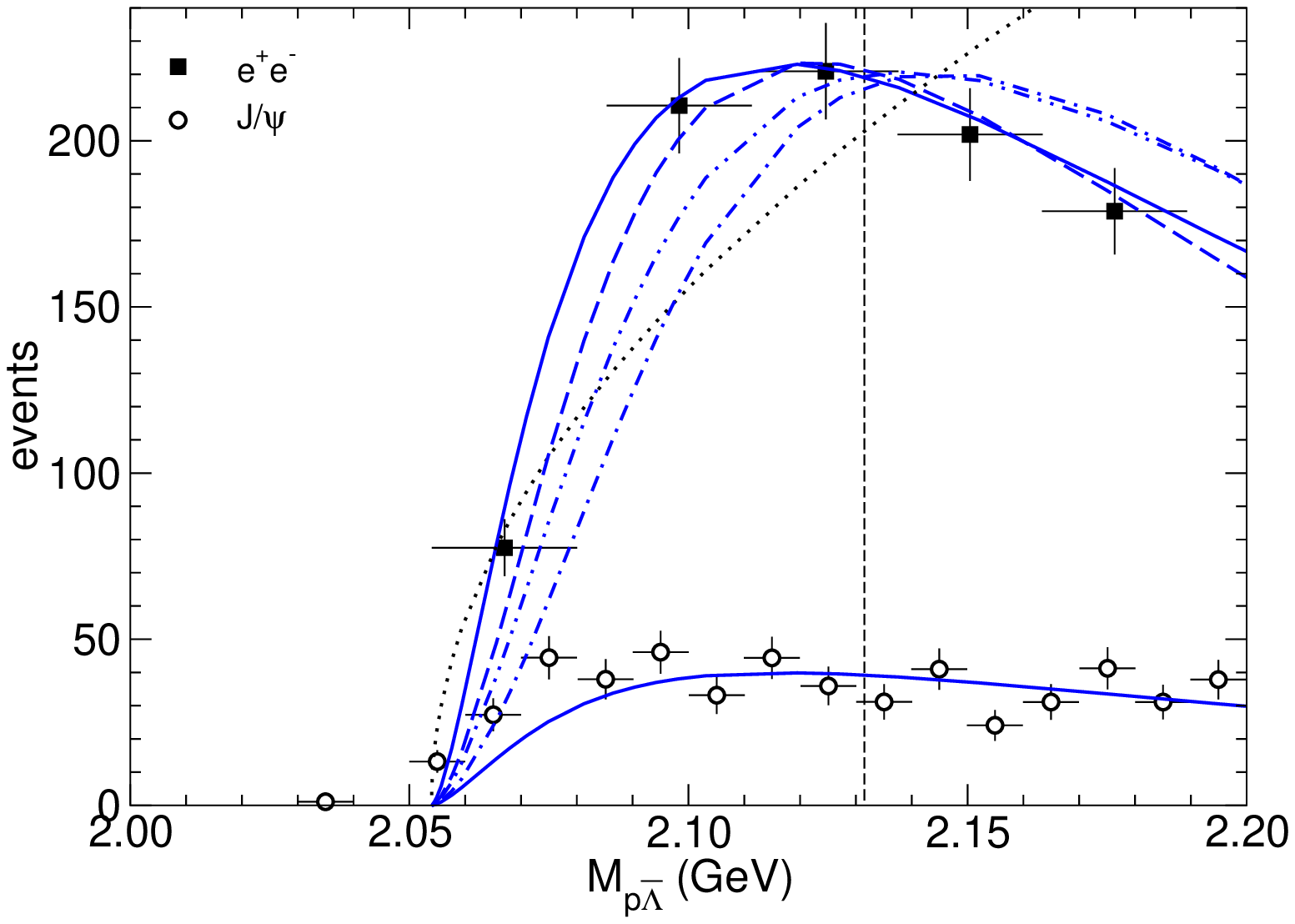}
\caption{Invariant-mass spectra for $\eeb\to K^-\pLb$ (open squares)
\cite{BESIII:2023} 
and $J/\psi \to K^-\pLb$ (black) \cite{BES:2004} (filled circles). 
The bin width for the former is $26$~MeV, while for the latter
it is $10$~MeV. 
Blue curves are based on the momentum dependence predicted by the $\LLb$ 
interaction I-IV \cite{Haidenbauer:2023} in the $^3P_1$ partial wave, 
corresponding to the quantum number $J^P=1^+$ of the 
resonance proposed by BESIII. The phase-space behavior is indicated by the
dotted line. The vertical dashed line marks the $\pLb$ threshold. 
}
\label{fig:Lpe}
\end{center}
\end{figure}

\begin{figure}
\begin{center}
\includegraphics[height=88mm,angle=-90]{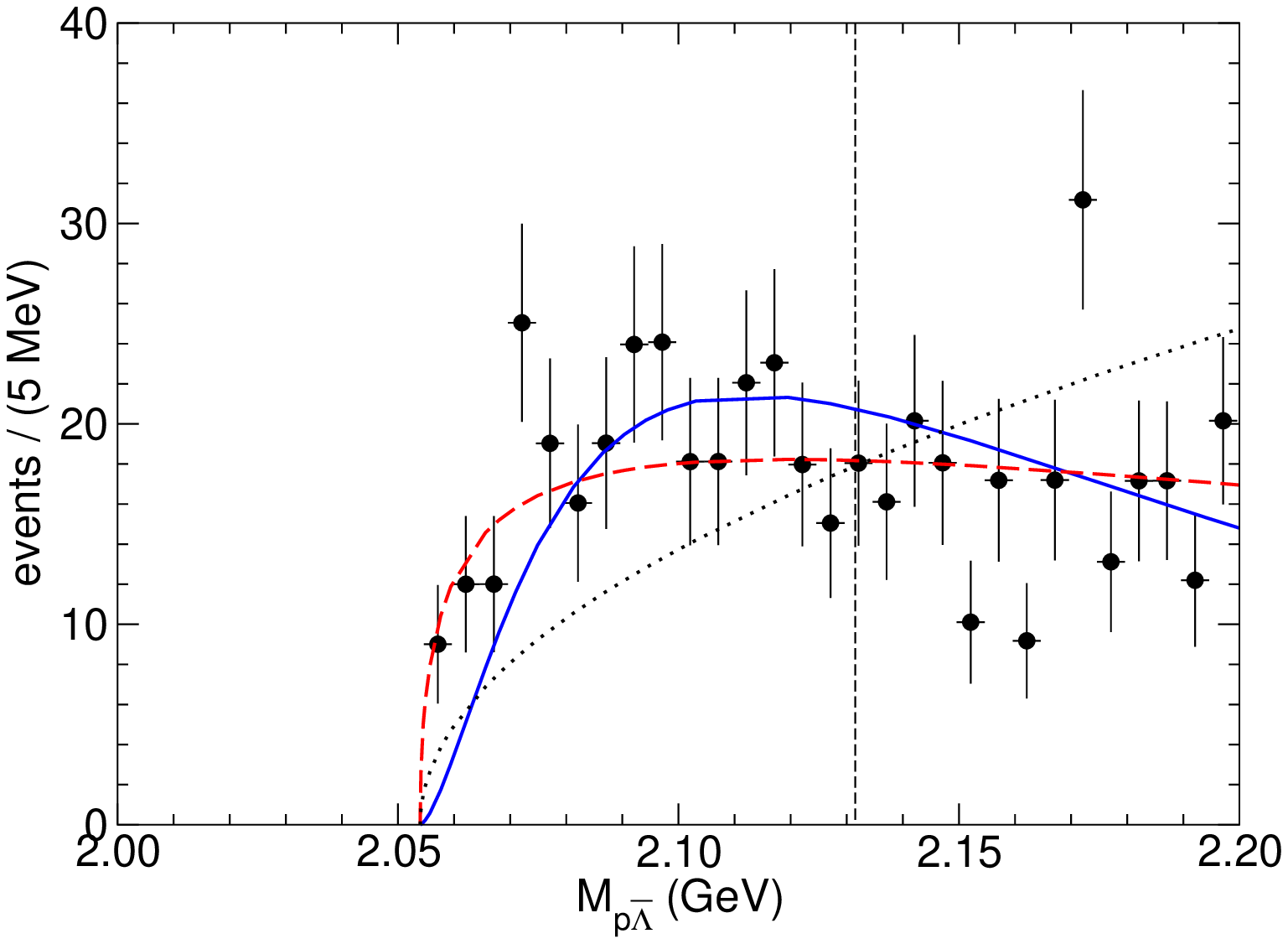}
\caption{Invariant-mass spectrum for $J/\psi \to K^-\pLb$  
\cite{BES:2004}. 
The solid (blue) curve is based on the $\LLb$ interaction~I
\cite{Haidenbauer:2023} in the $^3P_1$ 
partial wave while the dashed (red) curve is based on the $^3S_1$ partial wave.
}
\label{fig:Lpj}
\end{center}
\end{figure}


\subsection{The reactions $\eeb\to K^-\pLb$ and $J/\psi \to K^-\pLb$} 

In a recent paper the BESIII Collaboration reported evidence for a narrow
structure in the $\pLb$ system near threshold from a measurement of
$\eeb\to K^-\pLb$ \cite{BESIII:2023}. The spin and parity of the 
structure was determined to be $J^P = 1^+$ and the values
$m= 2084^{+4}_{-2} \pm 9$~MeV, $\Gamma = 58^{+4}_{-3}\pm 25$~MeV 
for its pole position were extracted from a fit to the line shape
with a relativistic Breit-Wigner function. We point out that doing this so
close to threshold is not appropriate. Already in 
2004 the BES Collaboration had measured the reaction 
$J/\psi \to K^-\pLb$ \cite{BES:2004} and an enhancement of the
near-threshold $\pLb$ invariant mass had been detected which could
be fitted by a resonance with parameters 
$m= 2075\pm 12 \pm 5$~MeV, $\Gamma = 90\pm 35$~MeV (assuming an $S$-wave) or 
$m= 2044\pm 17$~MeV, $\Gamma = 20\pm 45$~MeV ($P$-wave).

\begin{figure}
\begin{center}
\includegraphics[height=88mm,angle=-90]{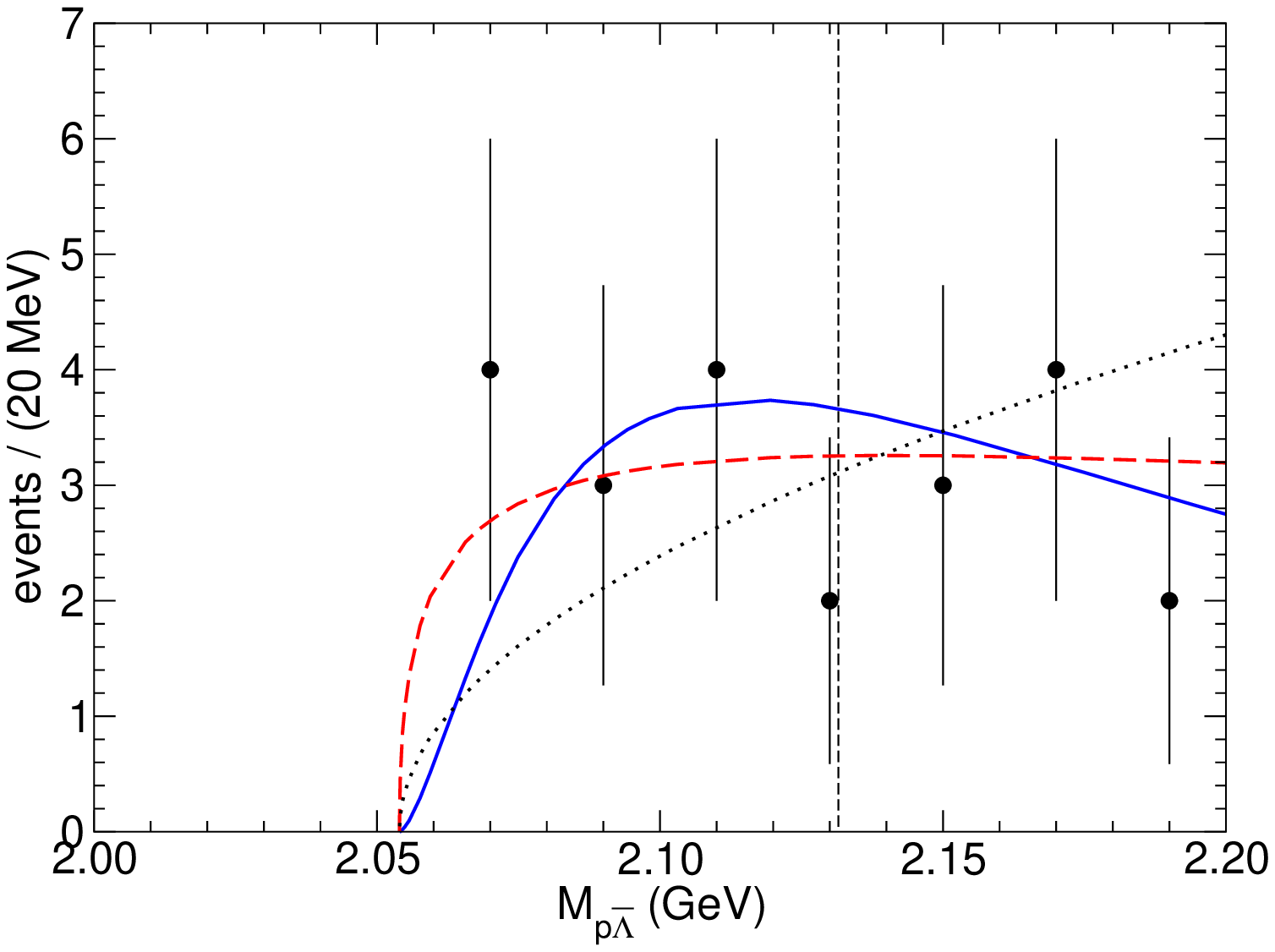}
\caption{Invariant-mass spectrum for $\psi (3686) \to K^-\pLb$  
\cite{BES:2004}.
Same description of curves as in Fig.~\ref{fig:Lpj}. 
}
\label{fig:Lpx}
\end{center}
\end{figure}

Since the $\pLb$ ($\pLb$) threshold is at $2053.95$~MeV it is obvious that 
those resonances overlap with the threshold region, considering their width. 
In such a case it is practically impossible to disentangle resonance effects
from an ordinary FSI. Therefore, it is tempting to explore whether a possible 
FSI in the $\pLb$ system could also explain the enhancement seen in the experiment. 
Assuming the spin and parity determined by BESIII, the $\pLb$ interaction should 
take place dominantly in the $^3P_1$ (or $^1P_1$) partial wave. 
Corresponding results are shown in Fig.~\ref{fig:Lpe} 
based on the momentum dependence of the $\LLb$ models I-IV 
\cite{Haidenbauer:1992,Haidenbauer:1992B} employed in our study of
the $\LLb$ FSI \cite{Haidenbauer:2023}. 
In order to guide the eye we show the phase space behavior (dotted line),
arbitrarily normalized to the data at an excess energy of $\sim 90$~MeV.
Furthermore we indicate the $p\bar\Sigma$ threshold by a vertical line. 
Actually, neither the $\pLb$ ($\pLb$) correlation functions 
nor the considered data on the $\pLb$ ($\pLb$) spectra do show 
any convincing evidence for the opening of the $p\bar \Sigma $ ($\bar p \Sigma$) 
threshold which is around $M_{\pLb} = 2.131$~GeV. 
This is quite different from the $p\Lambda$ system
where a sizable cusp at the $p \Sigma$ has been seen in many
experiments~\cite{Machner:2013,ALICE:2021}. We see that as further justification for the simplified
treatment of the $\pLb$ ($\pLb$) interaction in the present work. 

Obviously, and not unexpectedly, there is a sizable model dependence of the
predicted invariant-mass spectrum. But all results show the same qualitative
trend, namely an enhancement of the invariant mass near threshold.  
Moreover, intriguingly two of the model yield results that are quite well in 
line with the 
experiment. Specifically, model~I (solid line) reproduces the near-threshold 
behavior of the $\pLb$ invariant mass from the BESIII experiment 
$\eeb\to K^-\pLb$ \cite{BESIII:2023} (open squares) more or less exactly. 
Since the momentum dependence predicted by that model is so clearly favored 
by the data we select it as basis for the subsequent discussion. 

In the same figure, one can find also results for the $\pLb$ spectrum 
deduced from the reaction $J/\psi \to K^-\pLb$ \cite{BES:2004} (filled circles), 
again confronted with theory. With regard to the latter we show only the prediction 
based on model I for reasons of clarity. Also in this case the FSI reproduces the 
overall trend of the invariant-mass spectrum quite well, though there might
be a discrepancy very close to threshold. 
In order to shed light on that region, in Fig.~\ref{fig:Lpj} we compare our 
results with $\pLb$ data with refined binning, cf. Fig.~2(c) in \cite{BES:2004}.
In addition, we include predictions based on the $^3S_1$ partial wave of 
model~I (dashed red line). Obviously both scenarios are well in line with the
experiment, given the present uncertainties. 

Finally, in Ref.~\cite{BES:2004} results for the $\pLb$ invariant-mass spectrum 
from $\psi (3686) \to K^-\pLb$ were presented. Those are shown in Fig.~\ref{fig:Lpx},
again in comparison to predictions. For that reaction the statistics is much lower.
But still there is clear evidence for a deviation from the phase-space behavior 
\cite{BES:2004} and strong support for the presence of a FSI, which could be
either in the $^3P_1$ or $^3S_1$. 

\begin{figure}
\begin{center} 
\includegraphics[height=88mm,angle=-90]{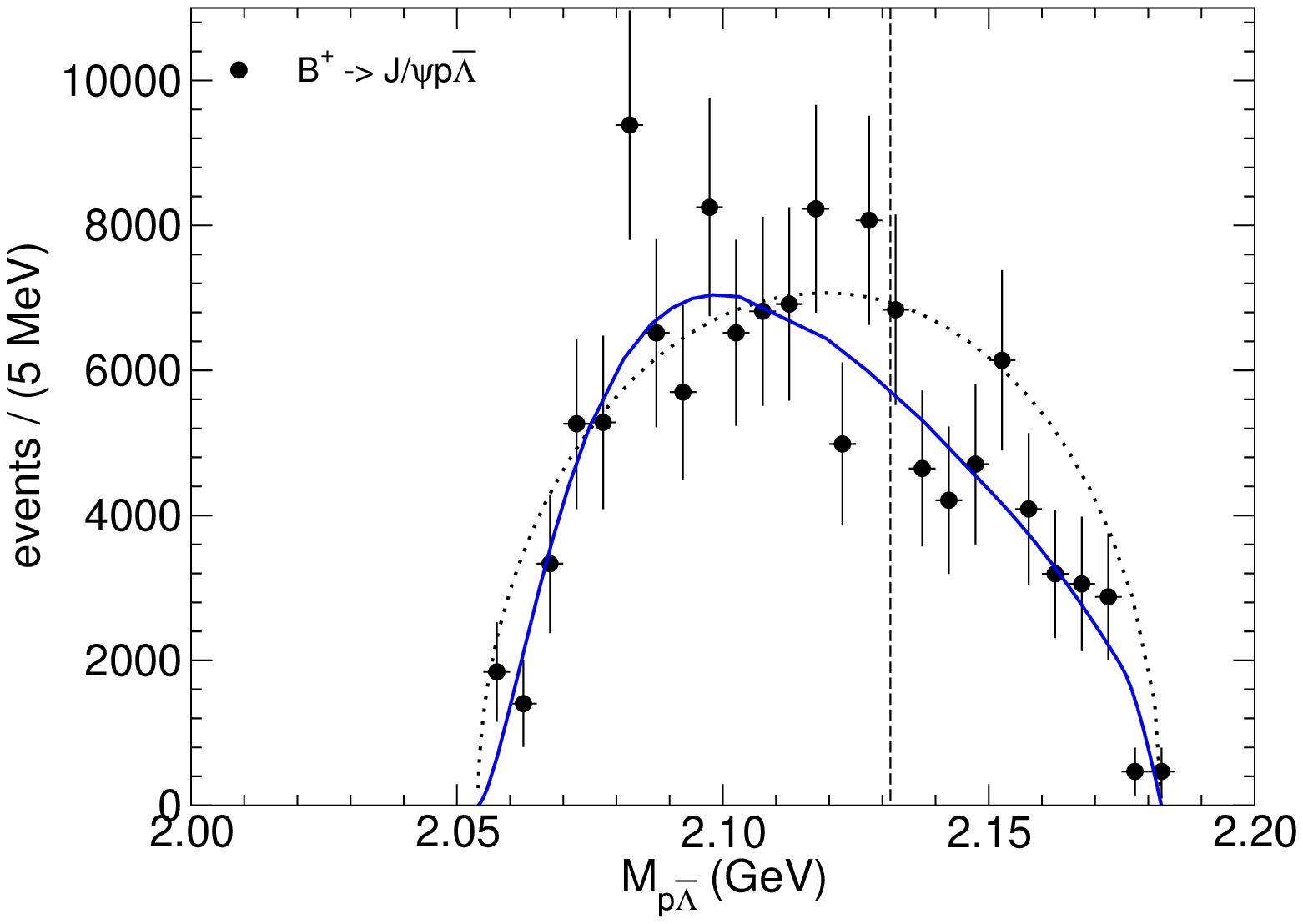}
\caption{Invariant-mass spectrum for $B^+ \to J/\psi\, \pLb$ from CMS 
\cite{CMS:2019},
The blue curve is based on the $\LLb$ interaction in the $^3P_1$ partial wave.
Same description of curves as in Fig.~\ref{fig:Lpj}. 
}
\label{fig:LpB}
\end{center}
\end{figure}

\begin{figure}
\begin{center}
\includegraphics[height=88mm,angle=-90]{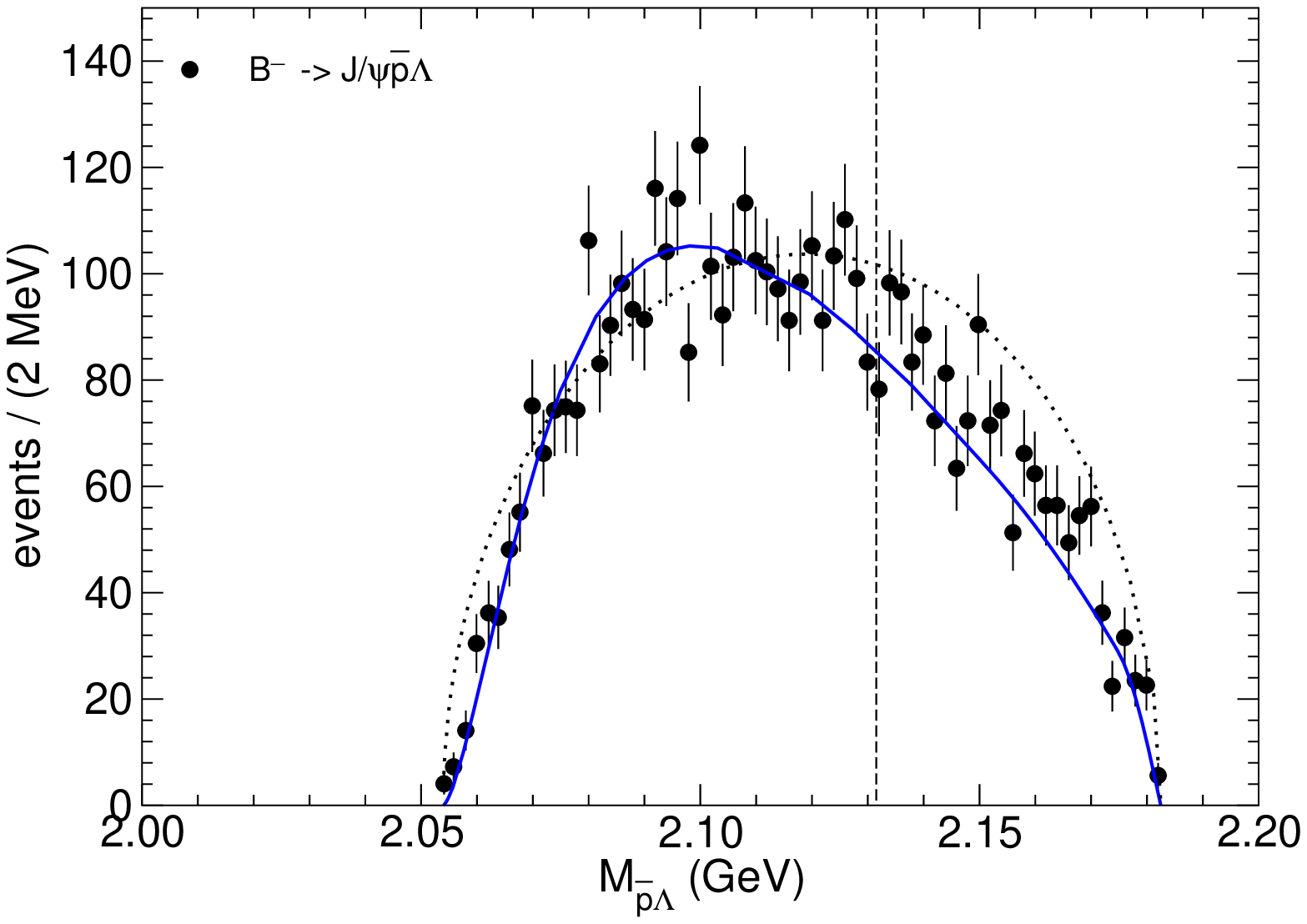}
\caption{Invariant-mass spectrum for $B^- \to J/\psi\, \bar p \Lambda$ from LHCb 
\cite{LHCb:2023}, 
The blue curve is based on the $\LLb$ interaction in the $^3P_1$ partial wave.
Same description of curves as in Fig.~\ref{fig:Lpj}. 
}
\label{fig:LpBM}
\end{center}
\end{figure}


\subsection{The reactions $B^+\to J/\psi \pLb $ and $B^- \to J/\psi \Lpb$}

In Figs.~\ref{fig:LpB} and \ref{fig:LpBM} we present the $\pLb$ ($\Lpb$)
invariant-mass spectrum measured in the reactions 
$B^+\to J/\psi \pLb $ and $B^- \to J/\psi \Lpb$ by the 
CMS \cite{CMS:2019} and LHCb \cite{LHCb:2023} Collaborations, respectively.
Those data are very interesting because of the excellent invariant-mass
resolution. However, there is also a caveat. The available phase space is
with less than $130$~MeV extremely small so that the interactions in the
other two-body final states ($J/\psi p$, $J/\psi \bar\Lambda$ or
$J/\psi \bar p$, $J/\psi \Lambda$) could already
distort the signal of the $\pLb $ ($\Lpb$) interaction and then 
Eqs.~(\ref{eq:dwba2}) are no longer applicable. 
One should keep that in mind. In any case, the region for realistic conclusions 
is certainly restricted to, say, $20$ to $30$~MeV from the threshold. Given
that restrictions it is really remarkable that the data follow very closely 
the predictions based on the $^3P_1$ partial wave near the threshold,
see Figs.~\ref{fig:LpB},~\ref{fig:LpBM}. We see that as clear signature
that the $1^+$ dominates here, and not the $1^-$ \cite{LHCb:2023}. 
Interestingly, with the $\pLb$ ($\Lpb$) interaction in the $^3P_1$ the
invariant mass over the whole phase space is described rather well, 
when the phase-space factor Eq.~(\ref{eq:INV2}) is appropriately taken into
account.

\begin{figure}
\begin{center}
\includegraphics[height=88mm,angle=-90]{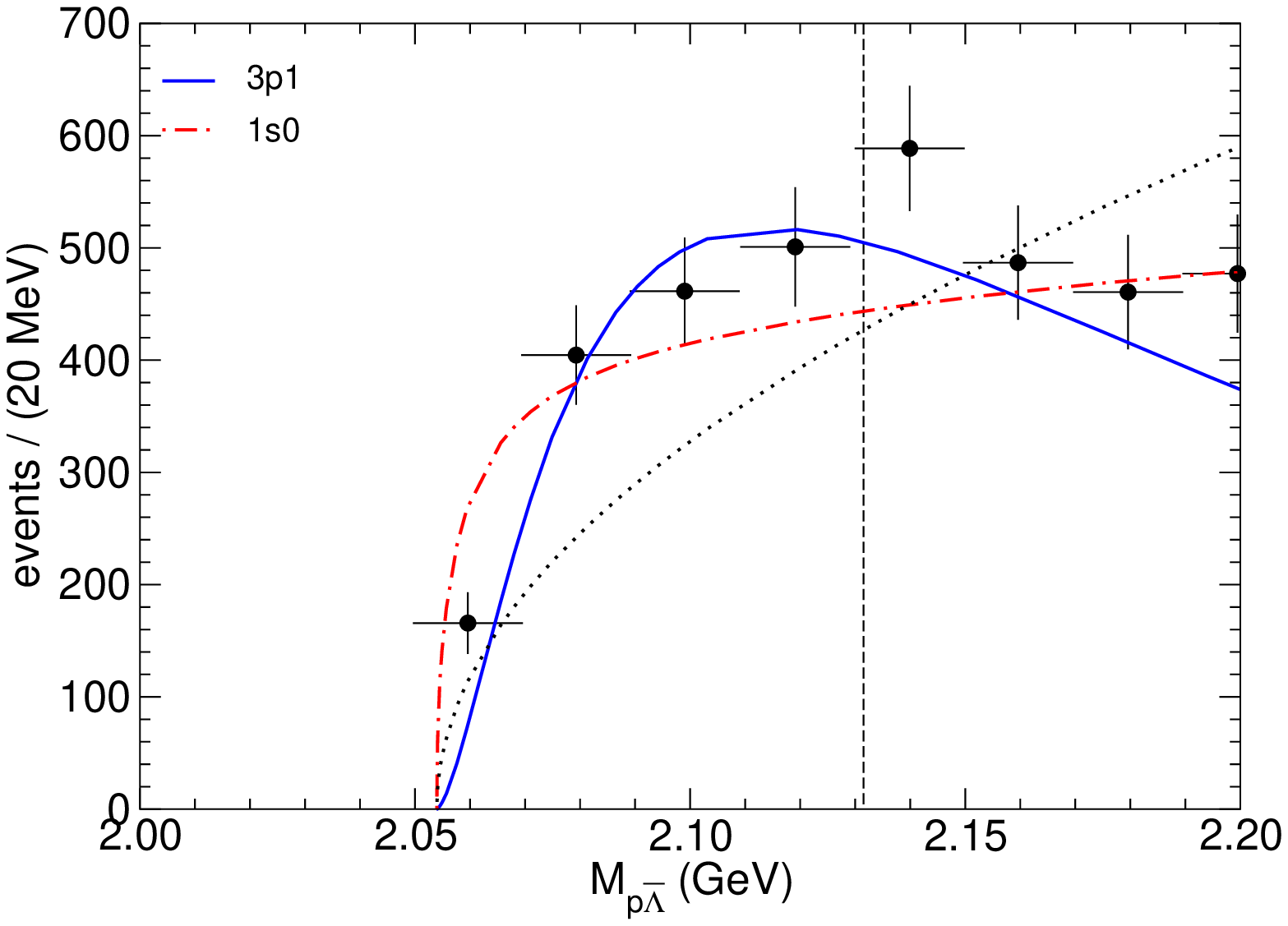}
\caption{Invariant-mass spectrum for $\chi_{c0} \to K^+\bar p\La$  
\cite{BESIII:2013}.
The solid (blue) curve is based on the $\LLb$ interaction in the $^3P_1$ 
partial wave while the dashed-dotted (red) curve is based on the $^1S_0$ 
partial wave.
}
\label{fig:Lpc}
\end{center}
\end{figure}

\begin{figure}
\begin{center}
\includegraphics[height=88mm,angle=-90]{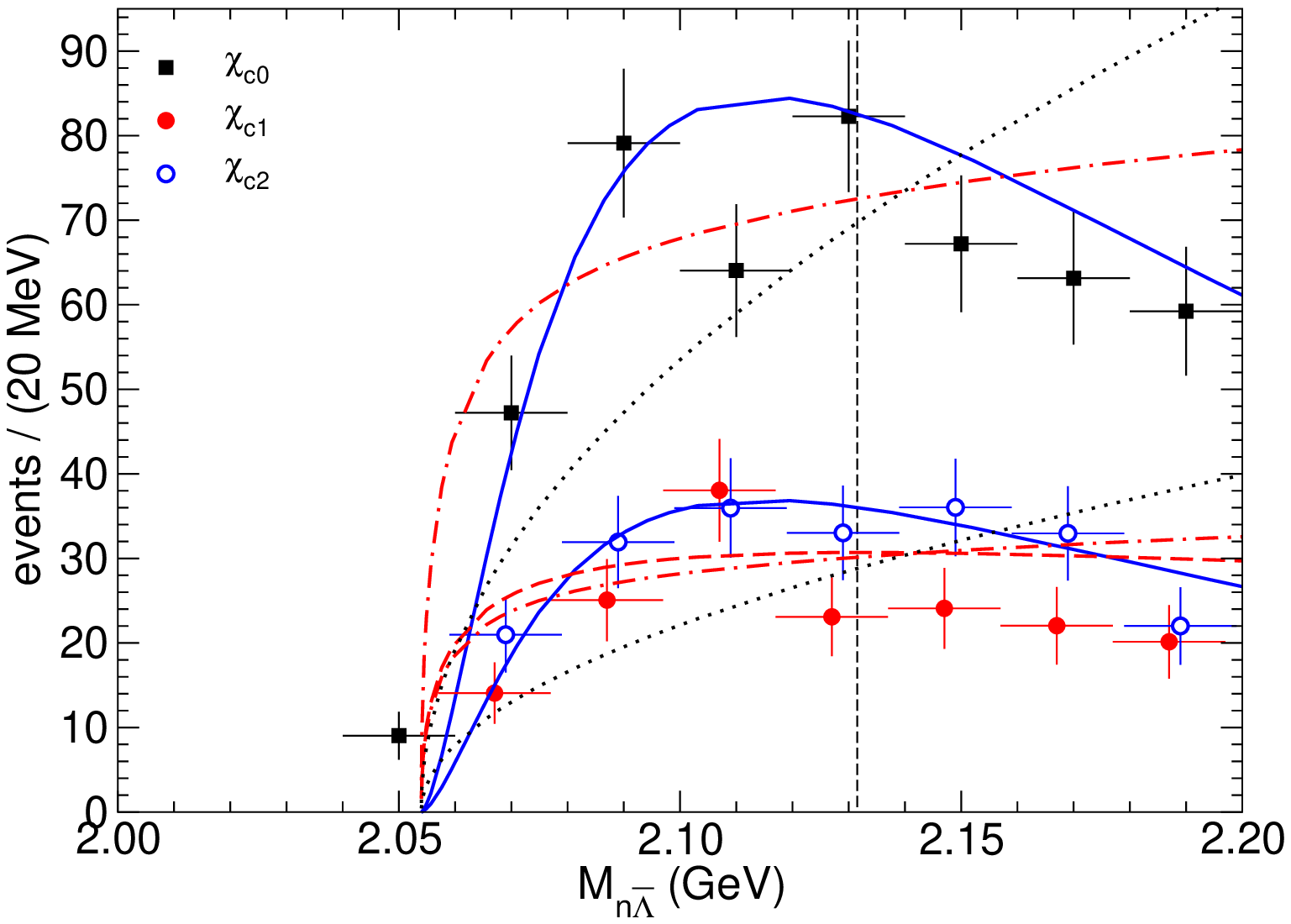}
\caption{Invariant-mass spectrum for $\chi_{cJ} \to K^{0}_S\bar n\La$  
($J=0,1,2)$ \cite{BESIII:2021c}. 
Same description of curves as in Figs.~\ref{fig:Lpj} and \ref{fig:Lpc}. 
}
\label{fig:Lnc1}
\end{center}
\end{figure}

\begin{figure}
\begin{center}
\includegraphics[height=88mm,angle=-90]{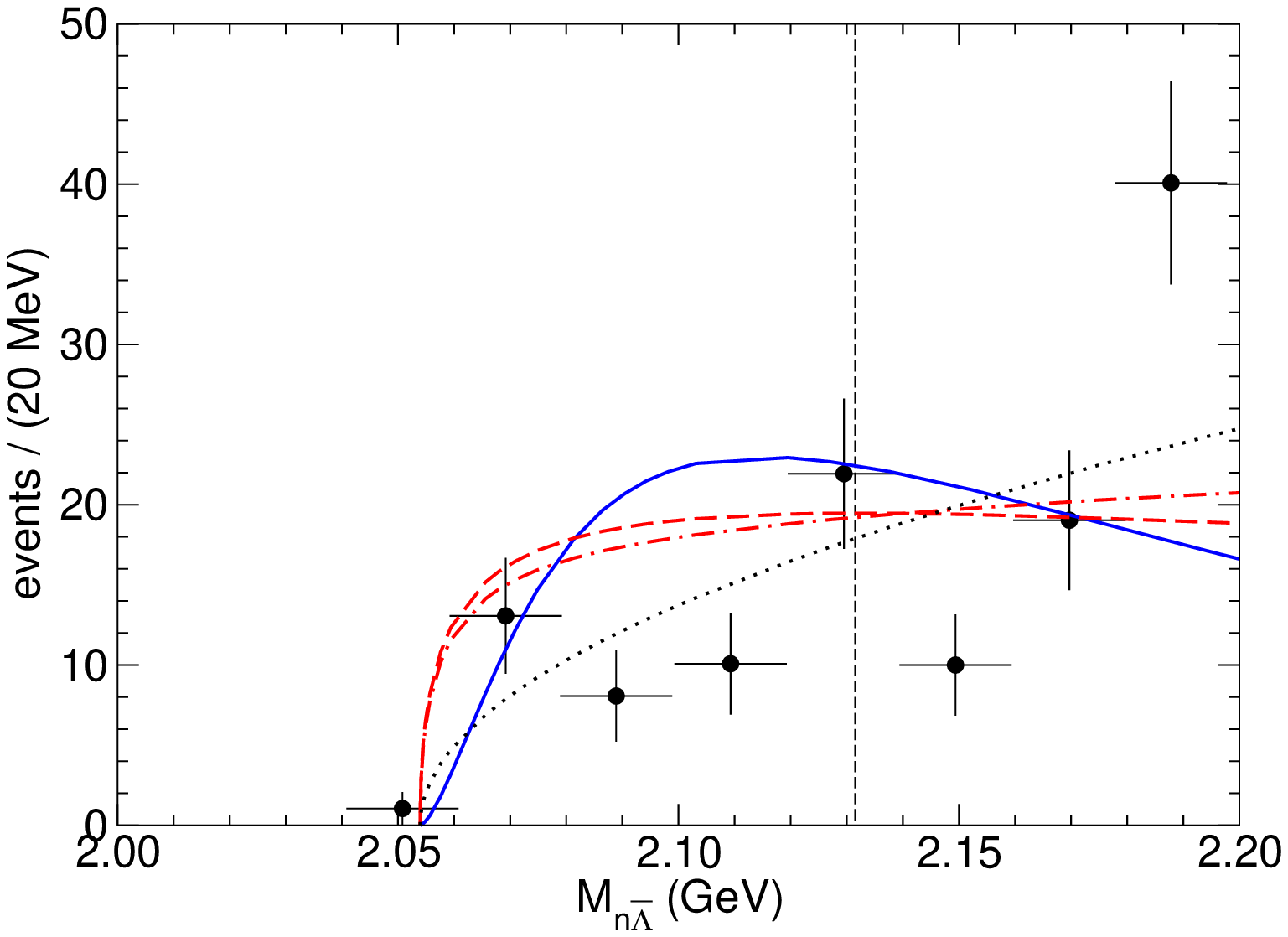}
\caption{Invariant-mass spectrum for $J/\psi \to K^{0}_S\bar n\La$  
\cite{BES:2007}. 
Same description of curves as in Figs.~\ref{fig:Lpj} and \ref{fig:Lpc}. 
}
\label{fig:Lnj}
\end{center}
\end{figure}

\begin{figure}
\begin{center}
\includegraphics[height=88mm,angle=-90]{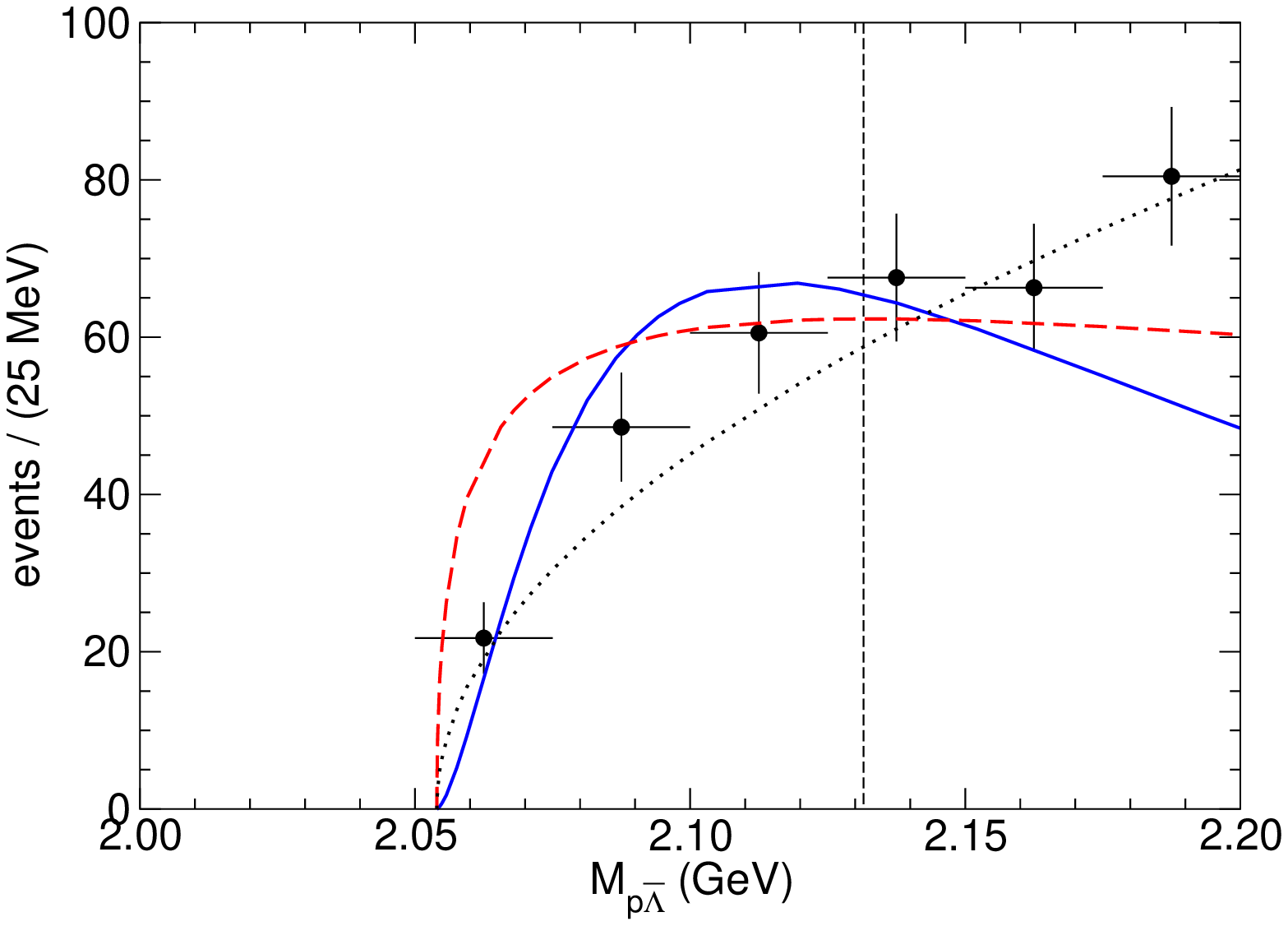}
\caption{Invariant-mass spectrum for $\psi (3686) \to K^{*+}\bar p\La$  
\cite{BESIII:2019}.
Same description of curves as in Fig.~\ref{fig:Lpj}. 
}
\label{fig:Lpks}
\end{center}
\end{figure}

\begin{figure}
\begin{center}
\includegraphics[height=88mm,angle=-90]{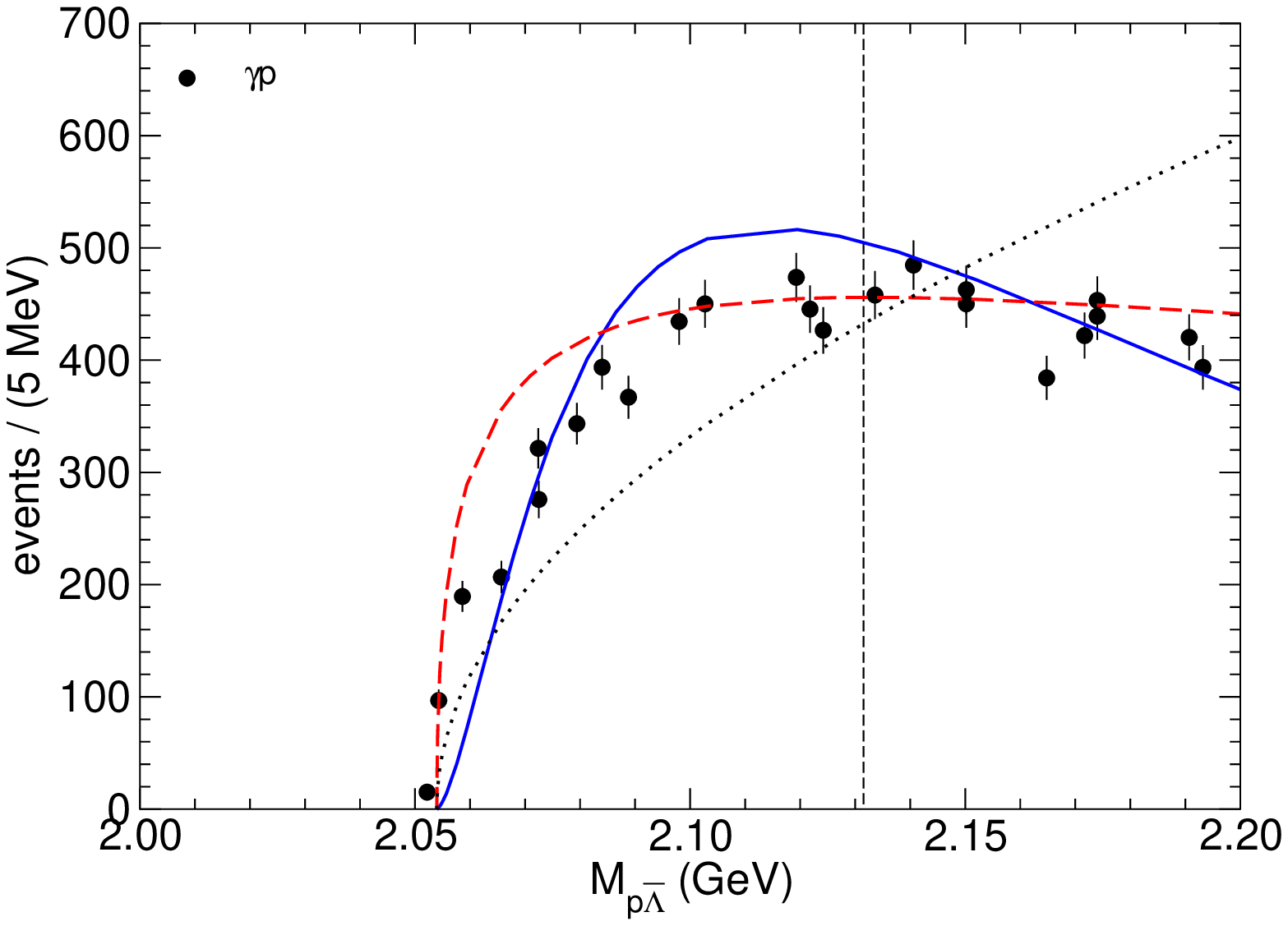}
\caption{Invariant-mass spectrum for $\gamma p \to \La\bar\La p$.
The red squares are preliminary data 
from a GlueX experiment \cite{Li:2020}. 
Same description of curves as in Fig.~\ref{fig:Lpj}. 
}
\label{fig:Lpgl}
\end{center}
\end{figure}


\subsection{The $\pLb$ invariant-mass spectrum in other reactions}

In Figs.~\ref{fig:Lpc} - \ref{fig:Lpgl} we show result for other reactions where 
the $\pLb$ invariant-mass spectrum has been measured. Also in those cases, 
in general we assume that either the $^3S_1$ or the $^3P_1$ partial waves provide 
the dominant FSI effect. An exception is the $\chi_{c0}$ decay, where the 
$^3S_1$ is not allowed, only the $^1S_0$, see Tab.~\ref{tab:JP}.

Let us start with $\chi_{cJ}$ decays where data are available from 
the reactions $\chi_{c0} \to K^+\bar p\La$ \cite{BESIII:2013} and
$\chi_{cJ} \to K^{0}_S\bar n\La$ ($J=0,1,2)$ \cite{BESIII:2021c} 
(Figs.~\ref{fig:Lpc} and \ref{fig:Lnc1}). In all cases the measured 
$N\bar \Lambda$ spectra are quite well in line with a FSI 
dominated by the $^3P_1$ partial wave.  
In this context let us mention that the BESIII Collaboration has
deduced a resonance in the $^1S_0$ ($0^-$) partial wave  
from those data on $\chi_{c0} \to K^+\bar p\La$ \cite{BESIII:2013} 
with the properties $m= 2053\pm 13$~MeV, $\Gamma = 292\pm 14$~MeV. 
In Ref.~\cite{Wang:2020wap} an explanation of the near-threshold enhancement 
is given in terms of a tree-level calculation that includes contributions from 
the intermediate $K(1830)$, $N(2300)$ and $\Lambda(1520)$ resonances. 
\\

The $n\bar\Lambda$ final state has been also measured 
in $J/\psi \to K^{0}_S\bar n\La$ \cite{BES:2007} (Fig.~\ref{fig:Lnj}). 
However, in this case conclusions are difficult to draw. 
The same applies to data on $\psi (3686)$ $\to K^{*+}\bar p\La$
\cite{BESIII:2019}, shown in Fig.~\ref{fig:Lpks}, which
are, in principle, even consistent with the phase-space behavior.
 
Finally, there is a measurement of the $\pLb$ invariant-mass spectrum 
from the reaction $\gamma p\to p\LLb$ by the GlueX Collaboration 
\cite{Li:2020}. As one can see from Fig.~\ref{fig:Lpgl}, those data 
are likewise roughly in line with our prediction based on the FSI in the 
$^3P_1$ partial wave. However, we want to emphasize that the results are 
still preliminary. Moreover, the small bin width in combination with 
the scale of the figure in Ref.~\cite{Li:2020} prevents a very reliable 
extraction of the actual values. 

Further measurements of the $\pLb$ ($\pLb$) invariant mass
have been reported for $B^0 \to \pi^- p \bar\La$ by the
Belle \cite{Belle:2003taw} and BaBar \cite{BaBar:2009ess}
Collaborations, but in these cases the bin width is $200$~MeV
or more so that possible near-threshold FSI effects cannot
be resolved. 
Therefore, we do not consider those data in the present work. 

\section{Summary and conclusions}

In the present work we have investigated invariant-mass spectra for the reactions 
$\eeb  \to K^-\pLb$ and $J/\psi \to K^-\pLb$ close to the $\pLb$ threshold.
Specific emphasis has been put on the effect of the interaction between the
final baryon-antibaryon pair which is taken into account rigorously. For it, 
as a working hypothesis and as guide line, a variety of $\LLb$ potential models 
have been utilized.
Those potentials, established in the analysis of data on the reaction 
$\ppbar\to \LLb$ from the LEAR facility at CERN, had been already
successfully applied in our analysis of FSI effects in the reactions 
$\eeb\to\LLb$, $\eeb \to \eta\LLb$, and $\eeb\to \phi\LLb$   
\cite{Haidenbauer:2023}. 
 
It turned out that the near-threshold invariant-mass dependence of the $\pLb$ 
spectra observed in those two reactions can be well reproduced by considering 
the $\pLb$ FSI. Specifically, the data for the 
reaction $\eeb  \to K^-\pLb$ can be perfectly described with an
FSI in the partial wave suggested by the amplitude analysis of the 
experiment. 
The high-precision measurements for the reactions $B^+ \to J/\psi\, p \bar\Lambda$  
and $B^- \to J/\psi\, \bar p \Lambda$ show also clear evidence for  FSI in
the $\pLb$ ($\Lpb$) $1^+$ state.  
Regarding the empirical information for other reactions our study is less conclusive, 
not least because in most cases the statistics of the experiments is significantly 
lower. 

As already concluded from other investigations in the past, there is strong evidence that 
the final-state interaction in baryon-antibaryon systems can and certainly does 
influence the properties of the invariant-mass spectrum in the near-threshold 
region. It has to be taken into account in any serious study that aims at a 
quantitative analysis of the threshold region.  If this is not done, any 
noticeable deviation from the pure phase-space behavior will be automatically 
attributed to near- or sub-threshold resonances. Allowing for the 
presence of FSI effects is the only way to avoid misinterpretations.  

\vspace{0.5cm}
\noindent
{\bf Acknowledgements:}
Work supported by the European Research Council (ERC) under the European
Uni\-on's Horizon 2020 research and innovation programme
(grant no.~101018170, EXOTIC), and by the DFG and the NSFC through
funds provided to the Sino-German CRC 110 ``Symmetries and
the Emergence of Structure in QCD'' (Project number 196253076 - TRR~110).
The work of UGM was also supported by the Chinese Academy of Sciences (CAS)
through a President's International Fellowship Initiative (PIFI)
(Grant No. 2018DM0034).


\end{document}